\newcommand{\dd}[1]{\hbox{\rm #1}}
\newcommand{\VEV}[1]{\left\langle {#1}\right\rangle} 
\newcommand{\beq}[1]{\begin{equation}  \label{#1} }
\newcommand{\ceq}{\end{equation}}
\newcommand{\bear}[1]{\begin{eqnarray}  \label{#1} }
\newcommand{\cear}{\end{eqnarray}}
\newcommand{\hide}[1]{} 
\def\I{\mbox{\boldmath $I$}}
\def\SS{\mbox{\boldmath $S$}}
\def\H{\mbox{\boldmath $H$}}
\def\B{\mbox{\boldmath $B$}}

\def\muB{\hbox{$\mu_B$}}

\def\musr{$\mu$SR}

\def\lcxy#1#2{La$_{0. #2}$Ca$_{0. #1}$\-MnO$_3$}
\def\laamo{La$_{1-x}$A$_x$\-MnO$_3$}

\def\lmo{LaMnO$_3$}

\def\lmod{LaMnO$_{3+\delta}$}
\def\lcmod{La$_{1-y}$Ca$_y$MnO$_{3+\delta}$}

\def\abs#1{\left| #1\right|}                
\documentstyle[epsfig,prl,aps]{revtex}
\begin{document}
\draft
\twocolumn
[\hsize\textwidth\columnwidth\hsize\csname @twocolumnfalse\endcsname
\title{Evidence of ultra-slow polaron dynamics in low-doped manganites 
from $^{139}$La NMR-NQR and muon spin rotation} 

\author{G. Allodi$^1$, M. Cestelli Guidi$^1$, R. De Renzi$^1$, A. Caneiro$^2$,
and L. Pinsard$^3$}
\address{$^1$Dipartimento di Fisica e 
Istituto Nazionale di Fisica della Materia, 
Universit\`a di Parma, I-43100 Parma, Italy \\
$^2$Centro Atomico Bariloche, San Carlos de Bariloche, Argentina\\
$^3$Laboratoire de Physico-Chimie des Solides, UMR 8648, Universit\'e 
Paris-Sud, 91405 Orsay C\'edex, France}

\date{\today}
\maketitle
\begin{abstract}
{
We report a $^{139}$La NMR investigation of low-doped insulating
manganite samples (\lmod\ and \lcmod) as a function of temperature.
A volume fraction with fast nuclear relaxations was revealed by the 
inhomogeneous loss of the NMR signal over a broad temperature interval.
Comparison with $\mu$SR data demonstrates that the 
{\it wipe out} of the $^{139}$La signal is mainly due to slowly fluctuating 
electric field gradients.
This provides strong evidence for the slow diffusion of lattice excitations,
identified with Jahn-Teller small polarons.
} 

\end{abstract} 
]  
\narrowtext

Lanthanum manganites \laamo\ (A = alkali-earth metal) have 
attracted great interest because of their correlated magnetic and 
transport properties, which include a colossal magnetoresistance (CMR) 
at suitable composition ($x \approx 0.3$). 
Recently the role of the
electron-lattice coupling has been emphasized. Millis {\it et al.}
\cite{millis} pointed out that the metal-insulator transition 
originating CMR 
arises from 
the cross-over from a large Jahn-Teller (JT) polaron to a small JT polaron 
regime. 
Experimental evidence 
of the involvement of lattice degrees of freedom in the electronic properties 
has been provided in recent years by a giant isotopic effect on $T_c$
 and on the EPR linewidth \cite{mueller}, by transport 
\cite{deteresa,ziese} and  electric thermopower \cite{jaime} measurements, and
by optical spectroscopies \cite{kim,ruani}.
The prominent role of the JT effect in manganites is now widely accepted 
among the physics community. Direct evidence 
of local lattice distortions, identified with JT polarons,
has also been provided by the pair distribution function (PDF) 
from neutron diffraction data \cite{billinge,louca}.
PDF, however, only yields the rms lattice distortion, without 
any information on lattice dynamics.

This paper addresses the issue of polaron dynamics in low doped, insulating
manganites by means of 
$^{139}$La NMR and muon spin rotation (\musr). 
We will show that, throughout this region of the phase diagram, $^{139}$La 
actually probes a slow diffusion of charged entities which 
are straightforwardly identified with lattice polarons.  

The investigated samples were \lmod\ and \lcmod , with hole
concentration $0 \le x \le 0.23$ ($x=y+2\delta$).
\lcxy{05}{95} was a twinned single crystal grown by the floating zone 
method. The other samples were sintered powders prepared by standard solid
state reaction. Details of sample preparation and characterization are 
reported elsewhere \cite{preparativa,hennion}. 
Resistivity of the doped samples was measured by the four-lead method. 
The most doped sample, La$_{0.8}$Ca$_{0.2}$MnO$_{3.015}$, exhibits a 
metallic behavior, with a resistivity maximum near $T_c$ = 190 K  
and $d\rho/d T >0$
at $T < T_c$, while the other samples are insulators.

$^{139}$La NMR was performed in an applied field of 7 T, in a temperature
range of 40-360 K. Spectra were recorded point by point with a standard 
$90^\circ - \tau - 90^\circ$ spin echo sequence on 
a phase coherent spectrometer. 
The pulse delay $\tau$ was kept as short as possible, 
limited by the dead time of the receiver following the transmission of a 
rf pulse (typically 10-14 $\mu$s).
At all temperatures the spin-spin relaxation rates $T_2^{-1}$ were determined
by varying $\tau$, and the echo amplitude decay was best fitted
by two exponentials.
The spin echo amplitudes were extrapolated back to $\tau = 0$, and 
divided by the NMR sensitivity $\propto \omega^2 /k_BT$. 
With this correction, the integrated amplitudes of each spectrum is 
proportional to the number of nuclei giving rise to the signal.
\musr\ spectra were recorded at ISIS, on either the MUSR or EMU
instruments. The data shown here are part of an extensive \musr\ investigation
of La manganites, reported elsewhere \cite{muoni}.
 
$^{139}$La NMR is sensitive both to the magnetic and to the electronic 
structure. Lanthanum nuclei ($\gamma/2\pi=6.014$ MHz/T) 
are coupled to the neighboring Mn electronic spins 
by a transferred hyperfine interaction. 
The local field $\B_{l}$, including the external field $\mu_0\H$, 
has the form  
\beq{eq:hyfi} 
\B_{l} \equiv \frac{\omega_L}{\gamma} = \frac{2\pi}{\gamma}
g\mu_B\sum_j^{n.n.}{\cal C}_j\VEV{\SS_j}+\mu_0 \H 
\ceq 
The hyperfine coupling constant 
is estimated in all lanthanum manganites as
${\cal C} \approx 0.1$ T/\muB \cite{gubkin,ca50,kumagai}.  
Owing to nearly cubic symmetry of the lanthanum site, the transferred 
hyperfine field on $^{139}$La is approximately proportional to the 
ferromagnetic moment
of the surrounding Mn octet. At $T \ll T_c$, the spontaneous field
 ranges from $\approx 3.5$ T ($\nu_L \approx 20$ MHz) in a fully 
ferromagnetic (F) environment \cite{gubkin,ca50}, down to a small but non zero
value of $\approx 300$~mT in the pure antiferromagnetic (AF) structure of 
\lmo \cite{kumagai}, due to the distortion of the perovskite cell.
In the paramagnetic (P) phase 
large frequency shifts 
proportional to the magnetic susceptibility
are produced by hyperfine couplings 
in the presence of external field.
\begin{figure}
\epsfig{figure=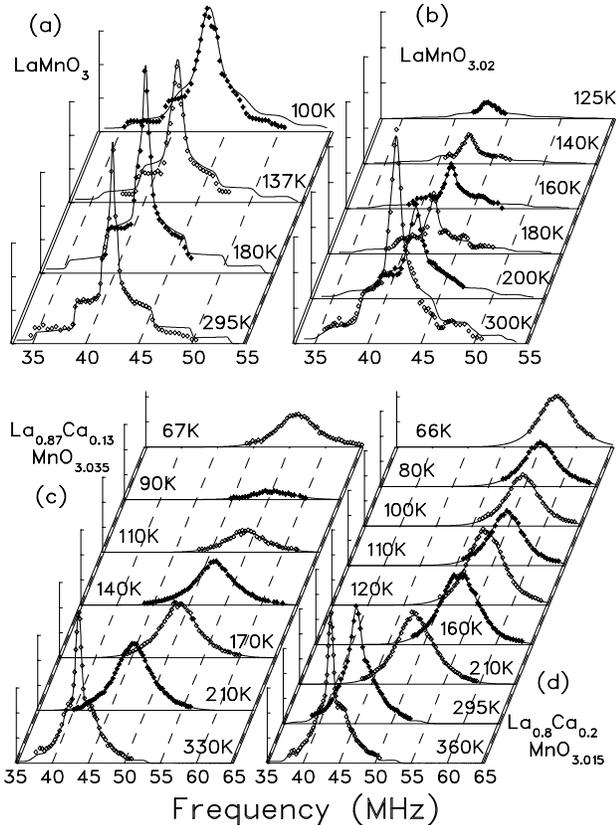,height=4.3in,width=3.2in,angle=0}
\caption{$^{139}$La NMR spectra in 7T 
($\gamma\mu_0 H/2\pi$ = 42.08 MHz) as a 
function of temperature from \lmo\ and LaMnO$_{3.02}$ (AF),  
La$_{0.87}$Ca$_{0.13}$MnO$_{3.035}$ (F insulator), and 
La$_{0.8}$Ca$_{0.2}$MnO$_{3.015}$ (F metal), respectively. }
\label{fig:allspectra}
\end{figure}
\noindent 
Note that in the present experiments we do not expect any critical
divergence of the nuclear relaxations rates near the magnetic transition,
since AF fluctuation modes yield negligible instantaneous fields at the
La site, and F fluctuations are suppressed by the intense external 
field \cite{moriya}.

In addition, $^{139}$La ($I=7/2$) is coupled to the local 
electric field gradient (EFG) tensor $V_{ij}$ through its electric 
quadrupole moment $Q$.
In the frame of reference of the EFG principal axes ($\abs{V_{zz}} 
\ge\abs{V_{yy}} \ge \abs{V_{xx}}$), the nuclear Hamiltonian accounting
for both magnetic and electric interactions is written as \cite{abragam}:
\beq{eq:zqham}
{\cal H}_{n} =\hbar\gamma\B_{l}\cdot\I + \frac{h\nu_Q}{6}[3I_z^2 -I(I\!+\!1) + \frac{1}{2}\eta(I_{+}^2\! +\! I_{-}^2)]
\ceq
Here $\nu_Q=3eQV_{zz}/[2hI(2I-1)]$, 
and $\eta=\abs{V_{xx}-V_{yy}}/V_{zz}$ is the EFG asymmetry parameter.  
The quadrupole interaction resolves the Zeeman transitions,
leading in a single crystal to angle-dependent multiline spectra with 
quadrupolar satellites.
In a polycrystalline sample the angular average of 
the satellite patterns gives rise to a characteristic powder spectrum.

The spectra of stoichiometric \lmo\ 
are plotted in Fig. \ref{fig:allspectra}a.
The spectra fit to quadrupolar powder patterns,  
\begin{figure}
\epsfig{figure=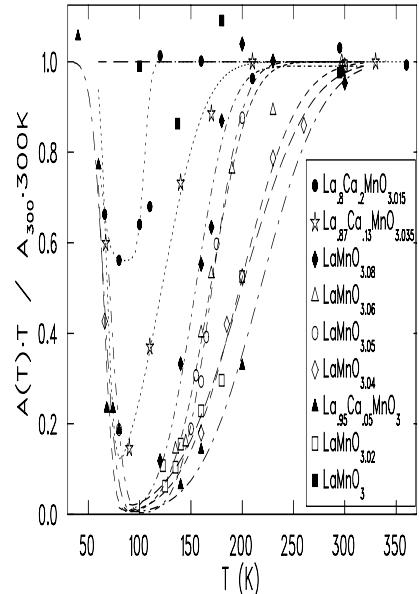,height=3.1in,width=2.1in,angle=90}
\caption{Relative $^{139}$La NMR integral amplitude from all the samples 
as a function of temperature. 
The amplitudes are corrected for $T_2$, $\nu_L^2$ and $T$.
The lines are best fits to the data
using the model described in the text.
}
\label{fig:A_T}
\end{figure}
\noindent 
broadened by magnetic interactions \cite{fitnuq}. 
The best fit parameters for the EFG are $\nu_Q=3.8(1)$ MHz, 
$\eta=0.94(2)$ at all temperatures, in 
close agreement with data reported by Kumagai et. al. \cite{kumagai}.
The paramagnetic shifts  
agree with magnetic susceptibility data, which follow a Curie-Weiss law 
for $S=2$, $g=2$ moments, and the magnetic line 
broadening is negligibly small at $T > T_N$.
This indicates a highly homogeneous and accurate 
stoichiometry for this sample.
The onset of an AF order is solely witnessed by a sizeable magnetic broadening
below $T_N = 139$ K. Figure \ref{fig:A_T} shows that 
the integrated spectral intensity, corrected for the NMR sensitivity,  
is constant within experimental accuracy, i.e. the signal originates from the 
whole sample volume at all temperatures.

Spectra from a subset of hole doped compounds are plotted in 
Fig. \ref{fig:allspectra}b-d. In the order phase they 
show magnetic shifts and line broadening increasing at increasing doping,
in agreement with a ferromagnetic moment increasing with $x$. 
The shifts at 70 K in LaMnO$_{3.04}$, LaMnO$_{3.06}$ (not shown, 
$T_c$ = 125 and 150 K) and 
La$_{0.87}$Ca$_{0.13}$MnO$_{3.035}$ ($T_c$ = 170 K)
are 7, 12 and 15 MHz, respectively. A 
comparison with the spontaneous frequency 
of $20$ MHz in the CMR ferromagnet \lcxy{3}{7}
indicates $\approx 35$\%, 60\% and 75\% ferromagnetically 
polarized Mn octets
in the three samples.
Furthermore, a dependence of the EFG on composition was observed.
In LaMnO$_{3.02}$ and \lcxy{05}{95} (both $T_N$ = 130 K)
the fit yields the same quadrupolar 
linewidth as in \lmo\ within experimental accuracy.
At doping concentrations $0.06 < x \le 0.1$, however, a progressively 
reduced EFG was detected, down to $\nu_Q=3.0$ MHz, $\eta = 0.3$ in 
LaMnO$_{3.05}$ ($T_c$ = 140 K), whereas no further EFG reduction was 
observed in 
the high temperature spectra at higher doping, up to $x=0.23$ in the metallic
La$_{0.8}$Ca$_{0.2}$MnO$_{3.015}$.
The EFG reduction is probably due 
to a lower static distortion of the MnO$_6$ octahedra in the doped compounds. 

\begin{figure}
\epsfig{figure=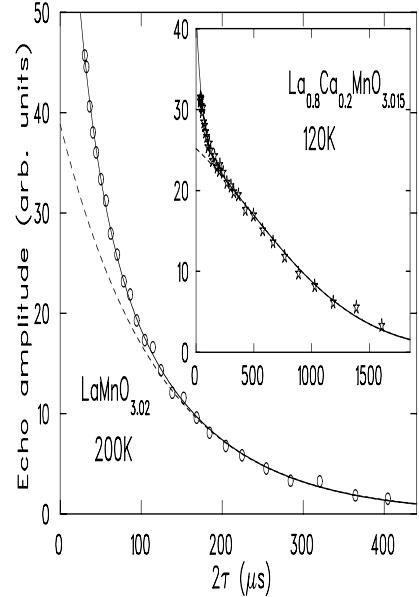,height=3.1in,width=2.1in,angle=90}
\caption{ 
$^{139}$La transverse depolarization curves ($T_2$ experiments) 
for LaMnO$_{3.02}$ at 200 K (main figure) 
and La$_{0.8}$Ca$_{0.2}$MnO$_{3.015}$ at 120 K (inset). 
The decays are best fitted (solid lines) with two time 
constants; the slower components are marked by the dashed lines.
}
\label{fig:t2demo}
\end{figure}

The most remarkable feature of the La spectra is however the 
strong reduction of the signal amplitude, which  occurs over
 a wide temperature range in all the doped {\em insulating} samples 
(Fig. \ref{fig:A_T}). 
In  LaMnO$_{3.04}$ ($T_c$ = 125 K) in particular, the signal was 
completely lost at $75\dd{ K} \le T \le 140\dd{ K}$.
The temperature interval where 
the signal diminishes depends on composition. It is maximum in the 
least doped compounds 
LaMnO$_{3.02}$ and \lcxy{05}{95}, where a missing fraction is already 
detected at $T\gg T_c$, right below room temperature. 
At increasing doping the
interval narrows, and its upper limit is lowered down to 140 K in 
La$_{0.87}$Ca$_{0.13}$MnO$_{3.035}$, 
a temperature well below $T_c$. 
The full signal amplitude is recovered in
all samples only at $T\le 60 K$. 
The missing amplitude in the insulating doped compounds 
is clearly due to fast 
and inhomogeneous spin-spin 
relaxations. This is confirmed by the relaxation time constants 
$T_2$ of the measured signals, which actually decrease down to 
values of order 10 $\mu$s
as temperature is lowered, and by echo amplitudes decaying with 
multi-exponential behavior, as shown in Fig. \ref{fig:t2demo}. 
The {\em wipe out} of the signal  
originates therefore from a fraction of nuclei with an even shorter
$T_2 \le 5$ $\mu$s, outside the time window of NMR because of the 
aforementioned dead time.
In the metallic sample La$_{0.8}$Ca$_{0.2}$MnO$_{3.015}$,
on the contrary,
the full signal amplitude is detected at all temperatures, except for 
a slight amplitude reduction at $60\dd{K} < T \le 110\dd{K}$. 
The small missing fraction in this sample is most likely due to a minority
insulating phase, either due to chemical inhomogeneity or to the electronic 
phase separation of hole-rich domains. This is 
indicated also by $^{139}$La nuclear relaxations, which 
reveal two distinct components (inset of Fig. \ref{fig:t2demo}).
The minority signal shows larger relaxation
rates increasing with decreasing temperature,
similar to those of the insulating compounds.
The majority component displays the opposite temperature dependence
typical of fully metallic manganites 
near optimum composition for CMR \cite{ca50,savosta}, and its
spin-spin relaxation eventually saturates  
to the gaussian behavior due to nuclear moments alone 
(homonuclear linewidth). 
We relate therefore the wipe out of the 
signal to the 
lightly doped insulating phase in all of our samples.

\begin{figure}
\epsfig{figure=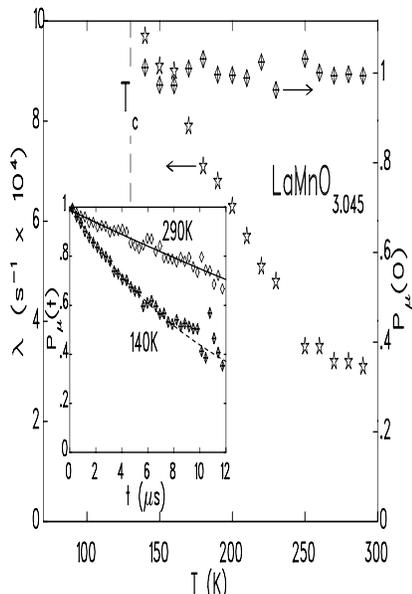,height=3.1in,width=2.1in,angle=90}
\caption{\musr\ of LaMnO$_{3.045}$ in a small longitudinal field of 20 G. 
Main figure: Lorentzian decay rate
$\lambda$ (stars) and initial spin polarization $P_\mu(0)$ (diamonds) as 
a function of temperature.
Inset: decay of the muon polarization as a function of time $t$ at two 
temperatures.
}
\label{fig:relaxmu}
\end{figure}

In principle, the fast nuclear relaxations responsible for the $^{139}$La 
wipe out might be due either to magnetic or to EFG fluctuations, 
$T_{2\mskip 2mu (La)}^{-1}=T_{2\mskip 2mu M}^{-1}+T_{2\mskip 2mu E}^{-1}$.
A clarification is provided by the comparison between NMR and \musr\ data. 
Implanted positive muons ($S_\mu=1/2$) are another local probe of magnetism, 
coupled only to the Mn electronic spins and not to the EFG.
Spontaneous precession frequencies $\nu_L^{\mu}\ge 80$ MHz have been 
observed in manganites at $T\ll T_{c,N}$, regardless the AF or F 
order \cite{heffner,muonilamno3}. 
Longitudinal muon relaxations in the P phase  
in a small applied field of 20 G 
are shown in Fig. \ref{fig:relaxmu}
for a typical low-doped sample (LaMnO$_{3.045}$, $T_c\approx130$K).
Muons exhibit the full initial 
polarization and comparatively low lorentzian 
relaxation rates $\lambda_\mu$ never exceeding $10^{5} \dd{s}^{-1}$ 
down to $T_c$. 
\noindent
Above $T=250K$ the 
ratio of the La and muon relaxation rates is 
$T_{2 \mskip 2mu (La)}^{-1}/\lambda_\mu=1/5$, which indicates that  
$T_{2\mskip 2mu M}^{-1}$ is at most $1/5$ of the muon rate.
This empirical upper limit implies that  
$T_{2\mskip 2mu M}\ge 5/\lambda_\mu \ge 50\,\mu$s also 
close to $T_c$
where the La signal vanishes.
If the 
La relaxation were dominated by the magnetic channel it would be safely 
within the time window of NMR, contrary to the experimental evidence. 
Note that relaxation rates scale as  
$T_{2\mskip 2mu M}^{-1}/\lambda_\mu = 16(\nu_{La}/\nu_\mu)^2$, 
where $\nu_{La}$ ($\nu_\mu$) is the instantaneous frequency in the La ($\mu$) 
fluctuating hyperfine field\cite{note}. Hence, even in the unrealistic worst 
case of fully F fluctuations, actually suppressed by the large external 
magnetic field, the spontaneous precession frequencies in the F state 
($\nu_L^{\mu}\ge80$ MHz, $\nu_L^{La} \le 20$ MHz) set a theoretical  limit 
$T_{2\mskip 2mu M} \ge \lambda_\mu^{-1}\ge 10$ $\mu$s, still 
within the time window of NMR. 
Therefore $T_{2\mskip 2mu E}^{-1}$ must dominate 
in the wiped out fraction. 

A similar loss of the Cu NQR signal intensity has been recently reported for 
the cuprate series La$_{2-x-y}$\-R$_y$\-Sr$_x$\-CuO$_4$ (R = rare earth), 
which exhibit stripe instability, and, close to the concentration $x= 1/8$,
the localization of static incommensurate 
charge-ordered 
stripes below a transition temperature 
$T_{charge}$ \cite{tranquada}. 
The wipe out of the Cu signal was ascribed to 
the {\em glassy} slowing down of the stripes, which are dynamic above 
$T_{charge}$\cite{hunt}. In the 
case of cuprates, however, 
it is still debated whether the diverging nuclear relaxations mainly involve 
spin or charge degrees of freedom \cite{curro}.

In lightly doped manganites 
no long range superstructure
has ever been detected above $T_c$.  Here, 
the wipe out of the La signal must 
arise from the diffusion of short range charge excitations coupled to 
lattice distortions.
A fine dispersion of these centers is actually indicated by 
the residual NMR signal, which also relaxes very fast and with several 
time constants, like in the case 
of a distribution of distances 
between the diffusing local distortion and the La nuclei.
Like in cuprates, however, 
the diffusion slows down continuously 
without any critical 
\noindent
behavior as temperature is lowered, as in a glass 
transition. 
A relaxation model adapted from Curro {\it et al.} \cite{curro} 
fits our $^{139}$La amplitude data (Fig. \ref{fig:A_T}); 
we assume
{\it i)} an instantaneous quadrupolar frequency $\Delta \omega_Q$
fluctuating with a lorentzian spectrum
and a correlation time $\tau_c = \tau_\infty \exp(E_a /k_BT)$, 
{\it ii)} an inhomogeneous $T_2^{-1} = \Delta\omega^2 \tau_c$
arising from a distribution of either 
the activation energy $E_a$ or 
$\Delta \omega_Q$, and
{\it iii)} the recovery of the signal in the static limit whereby 
$\tau_c$ is longer than the duration $\Delta t$ of the whole pulse sequence
($\Delta t \approx 30$ $\mu$s).
The narrowing of the wipe out temperature 
interval at increasing doping 
corresponds to the decrease of $\overline{E}_a/k_B$ from 750(50) K
in LaMnO$_{3.02}$ down to 350(50) K in La$_{0.87}$Ca$_{0.13}$MnO$_{3.035}$.
Assuming also for convenience EFG fluctuations of
comparable amplitude to
the static EFG (i.e., $\Delta \omega_Q/2\pi \approx 1$ MHz), 
the following threshold for the occurrence of the wipe out can be established: 
$\overline{\tau}_c \ge 10^{-9}$ s. We stress that direct determination 
of collective dynamics on this time scale is
only accessible to slow microscopic probes like NMR-NQR and \musr .

It is remarkable that the wipe out occurs throughout the whole
 {\em low doping insulating} region of the phase diagram, whereas it is 
absent in both the undoped and the metallic phases. 
This peculiar fact, namely the occurrence of this phenomenon under the 
combination of charge carriers and a macroscopic insulating behavior, 
strongly suggests that the diffusing centers are 
JT {\em small polarons}. 
The recovery of the full signal amplitude at approximately the same 
temperature $T_{rec} \approx 60$ K in all the samples denotes the 
freezing of the polarons,
which appear static to NMR at $T < T_{rec}$. 
It is worth noting that, in the same composition range investigated here,
Millis \cite{millis98} recently reported private communications by 
S.W. Cheong on a charge 
ordered (CO) phase at low temperature. 
The flat phase boundary of the CO region with $T_{CO}(x) \approx 60$-70 K is 
in good agreement with our nearly $x$-independent $T_{rec}$.
We therefore associate the polaron freezing with the onset of a CO state.
If the identification holds, the CO transition may be viewed as the transition
from a polaron liquid to a polaron crystal.

In conclusion, comparison of $^{139}$La NMR and \musr\ data from
low doped manganites
demonstrates the slow diffusion of charge-lattice excitations, which we
identify with small JT polarons.

This work was partially supported by a MURST-PRIN 2000 grant {\it ``Polaroni
magnetoelastici''}. 
Discussion with G. Guidi and M. Hennion
is gratefully acknowledged.

\end{document}